\documentstyle[12pt,axodraw]{article}
\newcommand{\JHEP}{J. High Energy Phys. }

\newcommand{\NP}{Nucl. Phys. }
\newcommand{\PR}{Phys. Rev. }

\newcommand{\PL}{Phys. Lett. }

\newcommand{\pspace}{\int d^3\mu}
\newcommand{\bfq}{{\bf q}}
\newcommand{\bfk}{{\bf k}}
\newcommand{\bfp}{{\bf p}}

\begin{document}
\baselineskip=20pt

\pagenumbering{arabic}

\vspace{1.0cm}
\begin{flushright}
LU-ITP 2002/009
\end{flushright}

\begin{center}
{\Large\sf Time-ordered perturbation theory on noncommutative spacetime: 
basic rules}\\[10pt]
\vspace{.5 cm}

{Yi Liao, Klaus Sibold}
\vspace{1.0ex}

{\small Institut f\"ur Theoretische Physik, Universit\"at Leipzig,
\\
Augustusplatz 10/11, D-04109 Leipzig, Germany\\}

\vspace{2.0ex}

{\bf Abstract}
\end{center}

Assuming the S-matrix on noncommutative (NC) spacetime can still be developped 
perturbatively in terms of the time-ordered exponential of the interaction 
Lagrangian, we investigate the perturbation theory of NC field theory. We first 
work out with care some typical Green functions starting from the usual concepts 
of time-ordering and commutation relations for free fields. The results are 
found to be very different from those in the naive approach pursued in the 
literature. A simle framework then appears naturally which can incorporate the 
new features of our results and which turns out to be the usual time-ordered
perturbation theory extended to the NC context. We provide the prescriptions 
for computing S-matrix elements and Green functions in this framework. We 
also emphasize that the naive seemingly covariant approach cannot be reproduced 
from the current one, in contrast to the field theory on ordinary spacetime. We 
attribute this to the phase-like nonlocal interaction intrinsic in NC field 
theory which modifies the analyticity properties of Green functions significantly. 

\begin{flushleft}


\end{flushleft}

\newpage
\section{Introduction}

The idea of describing coordinates in terms of noncommuting operators goes 
presumably back to Heisenberg and appeared already in a 
paper by Snyder $\cite{snyder}$ in 1947, extended by Yang $\cite{yang}$ in 
the same year. It was associated with problems of treating hadrons which are 
extended objects and the assumption that there might exist a fundamental 
length. At that time renormalization theory was not yet well developped, but 
considered as a `` distasteful procedure '' $\cite{snyder}$. Even today this 
judgement might be shared by some people. In any case the opinion is 
widespread that at least in a theory which describes consistently all 
phenomena down to the Planck length the notion of spacetime undergoes some 
drastic change. The attempts of Connes and Lott $\cite{lott}$ aiming at a 
reformulation of the standard model in terms of noncommutative (NC) geometry 
started a new era which has been continued in the context of string theory 
$\cite{string}$. An analysis which brought the subject neatly into the context 
of somewhat more conventional quantum field theory has been provided by 
Doplicher {\it et al} $\cite{doplicher}$. They showed that there exist 
representations of the coordinate operators such that for a class of minimal 
states in the state space of the system interactions can be written as the 
Moyal product of standard quantum fields as they have been introduced by Filk 
in another context $\cite{filk}$. The road is then open to perturbation theory 
which has as a new element just this non-local interaction.

Many calculations and model considerations have been performed in this vein 
and led to interesting new problems, most noticeably the ultraviolet-infrared 
mixing $\cite{mixing}$ and a potential violation of unitarity 
$\cite{unitarity}$. They were all based on the assumption that up to modifying 
vertices by NC phase factors the Feynman rules are the usual ones, in 
particular that lines are represented by the conventional Feynman propagators. 
Doubts on this have been raised recently in a paper by Bahns {\it et al}. 
$\cite{bahns}$ by reformulating in $\varphi^3$ theory the equation of motion in 
terms of the well-known Yang-Feldman equation $\cite{yfeldman}$. They arrived 
thereby at a solution which is manifestly Hermitian hence the theory must be 
unitary even if time-space NC is nonvanishing. This is, of course, in line 
with the general considerations of Ref. $\cite{doplicher}$, but does not yet 
explain why one arrives at violation of unitarity in the same model if one uses 
the Feynman rules of Filk $\cite{filk}$, as shown in Ref. $\cite{unitarity}$.
Our study shows that the answer is very simple. Starting from standard free 
field theory, i.e. the standard commutation relations and multi-particle Fock 
space, one still formally defines the S-matrix as 
$\displaystyle S=T\exp\left[i\int d^4x {\cal L}_{\rm int}\right]$. 
But when performing the contractions according to Wick's theorem properly one 
can never combine the contraction functions of positive and negative 
frequency to the causal Feynman propagator. This arises because when time does 
not commute with space, the time-ordering procedure does not commute with the 
star multiplication either. The naive approach in terms of Feynman propagators 
is thus not well founded. 

In the next section we shall illustrate this statement by a detailed analysis 
of some examples. The picture on how to proceed in the general case 
will appear naturally in section $3$. It turns out that the correct procedure 
of doing perturbation in NC field theory is just the time-ordered perturbation 
theory extended to the NC case. We shall spell out the prescriptions for 
computing S-matrix and amputated Green functions. We also provide an argument 
on how the seemingly covariant result in the naive approach cannot be 
reproduced from the time-ordered perturbation theory. We conclude in the last 
section.

\section{Analysis of some examples}

The basic quantity in quantum field theory is the Green function which 
determines the probability amplitude of S-matrix. We assume that perturbation 
theory for NC field theory can still be developped in terms of vacuum 
expectation values of time-ordered products of field operators so that the 
Green function is computed as a series expansion as usual,
\begin{equation}
\begin{array}{rcl}
G(x_1,\cdots,x_n)&=&\displaystyle 
<0|T\left(\varphi_1(x_1)\cdots\varphi_n(x_n)
\exp\left[i\int d^4x{\cal L}_{\rm int}\right]\right)|0>, 
\end{array}
\end{equation}
where ${\cal L}_{\rm int}$ is the interaction Lagrangian. When one expands 
the above in ${\cal L}_{\rm int}$, $T$ is meant to be taken before integration 
over spacetime is carried out. The time-ordering itself is also defined in the 
usual manner; for example, for free bosonic fields, it is 
\begin{equation}
\begin{array}{rcl}
T(\varphi_1(x_1)\varphi_2(x_2))&=&
\tau(x_1^0-x_2^0)\varphi_1(x_1)\varphi_2(x_2)+
\tau(x_2^0-x_1^0)\varphi_2(x_2)\varphi_1(x_1), 
\end{array}
\end{equation}
where $\tau$ is the step function. (We reserve $\theta$ for the NC parameter.)
In this section we shall work out explicitly some Green functions in NC field 
theory using the above definitions and free field commutators. The differences 
to the naive approach of NC field theory will be clearly illustrated by them.

\subsection{The vertex for the cubic scalar interaction}

Let us start with the following three-point function, 
\begin{equation}
\begin{array}{rcl}
G(x_1,x_2,x_3)&=&\displaystyle 
\int d^4x_4<0|T\left(\varphi(x_1)\varphi(x_2)\varphi(x_3)
(\varphi\star\varphi\star\varphi)(x_4)\right)|0>, 
\end{array}
\label{eq_0}
\end{equation}
which would occur in $\varphi^3$ theory. The star product is defined as 
\begin{equation}
\begin{array}{rcl}
(f_1\star f_2)(x)&=&\displaystyle 
\left[\exp\left(\frac{i}{2}
\theta^{\mu\nu}\partial^x_{\mu}\partial^y_{\nu}\right)
f_1(x)f_2(y)\right]_{y=x}, 
\end{array}
\end{equation}
where $x,y$ are the usual commutative coordinates and 
$\theta_{\mu\nu}$ 
is a real, antisymmetric, constant matrix characterizing the 
noncommutativity of spacetime, 
$[\hat{x}_{\mu},\hat{x}_{\nu}]=i\theta_{\mu\nu}$. 
As in the usual field theory, the 
basic idea to work out the above quantity is to reexpress the time-ordered 
product in terms of the normal-ordered product ($N$) plus contraction terms 
arising from field commutators when interchanging field operators. Sandwiched 
between the vacuum state, only the completely contracted terms will contribute. 
In the usual field theory this is accomplished by Wick's theorem and a 
completely contracted term is given by a product of Feynman propagators. The 
same result is taken for granted in the naive approach of NC field theory. 
But as 
we shall show below this is actually not the case when time does not commute 
with space. 

In the example of eq. $(\ref{eq_0})$, 
there are $4!$ time orders but we only have to consider two groups of them, 
namely, 
$x_1^0>x_2^0>(x_3^0~{\rm and}~x_4^0)$ and 
$(x_3^0~{\rm and}~x_4^0)>x_2^0>x_1^0$, 
while others may be obtained by interchange of coordinate indices. For both of 
them, the following result will be useful, 
\begin{equation}
\begin{array}{rl}
&
T(\varphi_3\varphi_4\star\varphi_4\star\varphi_4)\\
=&\displaystyle 
N(\varphi_3\varphi_4\star\varphi_4\star\varphi_4)\\
&\displaystyle 
+N\left(2D(0)\varphi_3\varphi_4+\pspace_{\bf{q}}\varphi_3\varphi(x_4-\tilde{q})
\right)\\
&\displaystyle 
+\tau_{34}N\left(D_{34}\star\varphi_4\star\varphi_4
                +\varphi_4\star D_{34}\star\varphi_4
		+\varphi_4\star\varphi_4\star D_{34}\right)\\
&\displaystyle 
+\tau_{43}N\left(D_{43}\star\varphi_4\star\varphi_4
                +\varphi_4\star D_{43}\star\varphi_4
		+\varphi_4\star\varphi_4\star D_{43}\right)\\
&\displaystyle 
+\tau_{34}\left(2D(0)D_{34}+\pspace_{\bf{q}}D(x_3-x_4+\tilde{q})\right)\\
&\displaystyle 
+\tau_{43}\left(2D(0)D_{43}+\pspace_{\bf{q}}D(x_4-x_3-\tilde{q})\right),
\end{array}
\label{eq_norder}
\end{equation}
where the star multiplication refers to $x_4$. 
Some explanations are in order. We have denoted the arguments of functions by 
the indices of their coordinates when no confusion arises; e.g., 
$\varphi_3=\varphi(x_3)$, $D_{34}=D(x_3-x_4)$ and 
$\tau_{43}=\tau(x_4^0-x_3^0)$. 
The above result is obtained by pushing annihilation (positive-frequency) 
operators to the right and creation (negative-frequency) operators to the 
left using the basic commutation relation between them, 
\begin{equation}
\begin{array}{rcl}
\varphi(x)&=&\varphi^+(x)+\varphi^-(x),\\
\left[\varphi^+(x),\varphi^-(y)\right]&=&\displaystyle D(x-y)\\ 
&=&\displaystyle 
\pspace_{\bf{p}}\exp[-ip_+\cdot (x-y)], 
\end{array}
\end{equation}
where $d^3\mu_{\bf{p}}=d^3{\bf p}[(2\pi)^3 2E_{\bf{p}}]^{-1}$  
is the standard phase space measure with 
$E_{\bf{p}}=\sqrt{{\bf p}^2+m^2}$, and 
$p_{\lambda}^{\mu}=(\lambda E_{\bf{p}},\bf{p})$ ($\lambda=\pm$) is the 
on-shell momentum with positive or negative energy. Due to NC of the star 
product we have to take care of the order of ordinary functions in addition 
to the order of annihilation and creation operators when interchanging 
operators. For example, 
\begin{equation}
\begin{array}{rl}
&N(\varphi_4\star\varphi_4\star\varphi_4)\\
=&
[\varphi^+\star\varphi^+\star\varphi^+
+\varphi^-\star\varphi^+\star\varphi^+\\
&\displaystyle 
+\varphi^-\star\varphi^-\star\varphi^+
+\varphi^-\star\varphi^-\star\varphi^-\\
&\displaystyle 
+(\varphi^-\underline{\star}\varphi^+)\star\varphi^+
+\varphi^-\underline{\star}(\varphi^+\star\varphi^+)\\
&\displaystyle 
+\varphi^-\star(\varphi^-\underline{\star}\varphi^+)
+(\varphi^-\star\varphi^-)\underline{\star}\varphi^+](x_4)\\
&\displaystyle 
+2D(0)\varphi_4+\pspace_{\bf q}\varphi(x_4-\tilde{q}),
\end{array}
\end{equation}
where $\tilde{q}_{\mu}=\theta_{\mu\nu}q^{\nu}$ and 
$\underline{\star}$ refers to the star product using $-\theta_{\mu\nu}$. 
The last but one term arises from contractions of the left and right fields 
with the middle one which is divergent as usual, 
while the last term comes from the contraction of the 
left with the right which is nonlocal because of the star product. These two 
terms are the origin of the third line and the last two lines in eq. 
$(\ref{eq_norder})$. These terms will finally contribute to the disconnected 
part of $G(x_1,x_2,x_3)$ and are thus ignored from now on. The second line in 
eq. $(\ref{eq_norder})$ does not contribute either when multiplied from the 
left by $\varphi_1\varphi_2$ or from the right by $\varphi_2\varphi_1$ and 
sandwiched between the vacuum. We are thus left with the fourth and fifth 
lines in eq. $(\ref{eq_norder})$. Before we proceed, we would like to stress 
that for NC time-space (i.e., $\theta_{0i}\ne 0$) it is not permitted to change 
the order of time-ordering and the star multiplication since the former 
involves distribution functions of time and the latter contains derivatives 
with respect to time. In this general case, we cannot move $\tau_{34}$ and 
$\tau_{43}$ inside of the star product to form with $D_{34}$ and $D_{43}$ the 
Feynman propagator $D_F(x_3-x_4)$. This is the main source of difference in 
the current approach from the conventional one pursued in the literature. Only 
when time commutes with space, the two approaches become identical. 

Let us go back to the computation of $G(x_1,x_2,x_3)$ and consider the 
time sequence $x_1^0>x_2^0>x_3^0>x_4^0$,
\begin{equation}
\begin{array}{rcl}
A&=&\displaystyle 
<0|\varphi_1\varphi_2
N\left(D_{34}\star\varphi_4\star\varphi_4
+\varphi_4\star D_{34}\star\varphi_4
+\varphi_4\star\varphi_4\star D_{34}\right)|0>.
\end{array}
\end{equation}
Up to disconnected terms, only the combination 
$\varphi_1^+\varphi_2^+\cdots\varphi_4^-\cdots\varphi_4^-\cdots$ 
contributes. Applying repeatedly, 
\begin{equation}
\begin{array}{rcl}
\varphi_1^+\varphi_2^+\varphi_4^-&=&
\varphi_4^-\varphi_1^+\varphi_2^+
+D_{24}\varphi_1^++D_{14}\varphi_2^+,
\end{array}
\end{equation}
we obtain, 
\begin{equation}
\begin{array}{rcl}
A&=&\{D_{14},D_{24},D_{34}\}_{\star}, 
\end{array}
\end{equation}
where the star refers only to $x_4$ and we have introduced the completely 
symmetrized sum of the star products, 
\begin{equation}
\begin{array}{rcl}
\{B_1,B_2,B_3\}_{\star}&=& 
\sum_{\pi_3}B_{\pi(1)}\star B_{\pi(2)}\star B_{\pi(3)}, 
\end{array}
\end{equation}
where $\pi_3$ is the permutation of three objects. 
The above result is symmetric in $x_1$ and $x_2$ and thus applies as well to 
the time sequence $x_2^0>x_1^0>x_3^0>x_4^0$. Including the case of 
$x_4^0>x_3^0$, we have, up to disconnected terms, 
\begin{equation}
\begin{array}{l}
<0|T\left(\varphi_1\varphi_2\varphi_3
(\varphi\star\varphi\star\varphi)(x_4)\right)|0>\\
=\tau_{34}\{D_{14},D_{24},D_{34}\}_{\star}
+\tau_{43}\{D_{14},D_{24},D_{43}\}_{\star}, 
{\rm ~for~}T_{12}T_{34},
\end{array}
\end{equation}
where $T_{12}T_{34}$ denotes the time sequence 
$(x_1^0{\rm ~and~}x_2^0)>(x_3^0{\rm ~and~}x_4^0)$.
The opposite case of $(x_3^0{\rm ~and~}x_4^0)>(x_1^0{\rm ~and~}x_2^0)$ is 
similarly computed to be, 
\begin{equation}
\begin{array}{rl}
T_{34}T_{12}:&\tau_{34}\{D_{41},D_{42},D_{34}\}_{\star}
+\tau_{43}\{D_{41},D_{42},D_{43}\}_{\star}. 
\end{array}
\end{equation}
The other two pairs of cases are obtained by permutation of indices $1,2,3$:
\begin{equation}
\begin{array}{rl}
T_{23}T_{14}:&\tau_{14}\{D_{24},D_{34},D_{14}\}_{\star}
+\tau_{41}\{D_{24},D_{34},D_{41}\}_{\star},\\
T_{14}T_{23}:&\tau_{14}\{D_{42},D_{43},D_{14}\}_{\star}
+\tau_{41}\{D_{42},D_{43},D_{41}\}_{\star},\\
T_{31}T_{24}:&\tau_{24}\{D_{34},D_{14},D_{24}\}_{\star}
+\tau_{42}\{D_{34},D_{14},D_{42}\}_{\star},\\
T_{24}T_{31}:&\tau_{24}\{D_{43},D_{41},D_{24}\}_{\star}
+\tau_{42}\{D_{43},D_{41},D_{42}\}_{\star}.
\end{array}
\end{equation}

The above results can be combined into a compact expression. Since the 
connected contribution involves exclusively functions 
$D_{j4}$ and $D_{4j}$ ($j=1,2,3$), 
we hope that they are also accompanied exclusively by step functions 
$\tau_{j4}$ and $\tau_{4j}$. 
This is indeed the case as in ordinary field theory. Actually all terms except 
$\{D_{14},D_{24},D_{34}\}_{\star}$ and $\{D_{41},D_{42},D_{43}\}_{\star}$
are already in the desired form. There are three contributions to 
$\{D_{14},D_{24},D_{34}\}_{\star}$ 
arising respectively from the time sequences 
$(x_1^0{\rm ~and~}x_2^0)>x_3^0>x_4^0$,
$(x_2^0{\rm ~and~}x_3^0)>x_1^0>x_4^0$ and 
$(x_3^0{\rm ~and~}x_1^0)>x_2^0>x_4^0$,
whose union is identical to the sequence 
$(x_1^0{\rm ~and~}x_2^0{\rm ~and~}x_3^0)>x_4^0$ 
corresponding to the product of step functions
$\tau_{14}\tau_{24}\tau_{34}$.
A similar combination occurs for $\{D_{41},D_{42},D_{43}\}_{\star}$,  so that 
\begin{equation}
\begin{array}{rl}
&<0|T\left(\varphi_1\varphi_2\varphi_3
(\varphi\star\varphi\star\varphi)(x_4)\right)|0>\\
=&\sum_{\lambda_1\lambda_2\lambda_3}
\tau_{14}^{\lambda_1}\tau_{24}^{\lambda_2}\tau_{34}^{\lambda_3}
\{D_{14}^{\lambda_1},D_{24}^{\lambda_2},D_{34}^{\lambda_3}\}_{\star}, 
\end{array}
\label{eq_a}
\end{equation}
where $\lambda_j=\pm$ defines the direction of time in the relevant function, 
\begin{equation}
\begin{array}{l}
\tau_{j4}^{\lambda_j}=\left\{
                        \begin{array}{rl}
                        \tau_{j4},{\rm ~for~}\lambda_j=+\\
                        \tau_{4j},{\rm ~for~}\lambda_j=-
			\end{array}
		      \right.,
D_{j4}^{\lambda_j}=\left\{
                   \begin{array}{rl}
		   D_{j4},{\rm ~for~}\lambda_j=+\\
		   D_{4j},{\rm ~for~}\lambda_j=-
		   \end{array}		      
		   \right..  
\end{array}
\end{equation}
We would like to emphasize again that it is generally not permitted to move 
the step functions into the star products in eq. $(\ref{eq_a})$. Only when 
time commutes with space, we are allowed to do so and the result of the naive 
approach is reproduced,  
\begin{equation}
\{D_F(x_1-x_4),D_F(x_2-x_4),D_F(x_3-x_4)\}_{\star}, 
{\rm ~for~}\theta_{0i}=0,
\end{equation}
where $D_F(x_j-x_4)=\tau_{j4}^+D_{j4}^++\tau_{j4}^-D_{j4}^-$ is the Feynman 
propagator in coordinate space. 

To transform into momentum space we first work out explicitly the quantity in 
eq. $(\ref{eq_a})$. Using the expressions, 
\begin{equation}
\begin{array}{rcl}
\tau_{j4}^{\lambda_j}&=&\displaystyle 
\frac{i\lambda_j}{2\pi}\int_{-\infty}^{\infty}ds_j
\frac{\exp[-is_j(x_j^0-x_4^0)]}{s_j+i\epsilon\lambda_j},\\
D_{j4}^{\lambda_j}&=&\displaystyle 
\pspace_{\bfp_j}\exp[-ip_{j\lambda_j}\cdot(x_j-x_4)],
\end{array}
\label{eq_rep}
\end{equation}
we have, for example, 
\begin{equation}
\begin{array}{rl}
&\tau_{14}^{\lambda_1}\tau_{24}^{\lambda_2}\tau_{34}^{\lambda_3}
\left(D_{14}^{\lambda_1}\star D_{24}^{\lambda_2}\star D_{34}^{\lambda_3}
\right) \\ 
=&\displaystyle 
\prod_{j=1}^3
\left[\frac{i\lambda_j}{2\pi}\int_{-\infty}^{\infty}ds_j\pspace_{\bfp_j}
\frac{\exp[-ip_{j}\cdot(x_j-x_4)]}{s_j+i\epsilon\lambda_j}\right]
\exp[-i(p_{1\lambda_1},p_{2\lambda_2},p_{3\lambda_3})]\\
=&\displaystyle 
\prod_{j=1}^3
\left[\int \frac{d^4p_j}{(2\pi)^4}iP_{\lambda_j}(p_j)\exp[-ip_j\cdot(x_j-x_4)]
\right]\exp[-i(p_{1\lambda_1},p_{2\lambda_2},p_{3\lambda_3})],
\end{array}
\label{eq_b}
\end{equation}
where we introduced $p_j^0=s_j+\lambda_j E_{\bfp_j}$ and 
\begin{equation}
\begin{array}{rcl}
P_{\lambda}(k)&=&\displaystyle 
\frac{\lambda}{2E_{\bfk}[k^0-\lambda(E_{\bfk}-i\epsilon)]}\\
&=&\displaystyle 
\left(2E_{\bfk}[\lambda k^0-(E_{\bfk}-i\epsilon)]\right)^{-1},\\
(k_1,k_2,\cdots,k_n)&=&\displaystyle 
\sum_{i<j}k_i\wedge k_j, 
\end{array}
\end{equation}
with $p\wedge q=1/2~\theta_{\mu\nu}p^{\mu}q^{\nu}$. The pair of 
parentheses introduced above has some properties to be used implicitly 
later on. It changes sign when the order of arguments is reversed. The 
shift of sign in some arguments is identical to the shift of sign in 
the remaining arguments. 
It is important to note that the wedge product in the NC phase of 
eq. $(\ref{eq_b})$ involves only on-shell momenta of positive ($\lambda_j=+$) 
or negative ($\lambda_j=-$) energy corresponding to propagation in the time 
direction of $x_j^0>x_4^0$ or $x_j^0<x_4^0$. Including all permutations of 
$D$ factors in eq. $(\ref{eq_a})$ which amounts to summing over permutations 
of the NC phase, and integrating over $x_4$, we arrive at, 
\begin{equation}
\begin{array}{rl}
&G(x_1,x_2,x_3)\\
=&\displaystyle 
\sum_{\lambda_1\lambda_2\lambda_3}\prod_{j=1}^3
\left[\int \frac{d^4p_j}{(2\pi)^4}iP_{\lambda_j}(p_j)\exp(-ip_j\cdot x_j)
\right]\\
&\displaystyle\times(2\pi)^4\delta^4(p_1+p_2+p_3)
\sum_{\pi_3}\exp[-i(p_{\pi(1)\lambda_{\pi(1)}},p_{\pi(2)\lambda_{\pi(2)}},
p_{\pi(3)\lambda_{\pi(3)}})]. 
\end{array}
\end{equation}
Transforming into momentum space is now straightforward, 
\begin{equation}
\begin{array}{rl}
&\hat{G}(k_1,k_2,k_3)\\ 
=&\displaystyle\prod_{j=1}^3\left[\int d^4x_j\exp(-ik_j\cdot x_j)\right]
G(x_1,x_2,x_3)\\
=&\displaystyle 
(2\pi)^4\delta^4(k_1+k_2+k_3)
\sum_{\lambda_1\lambda_2\lambda_3}\prod_{j=1}^3
\left[iP_{\lambda_j}(k_j)\right]\\
&\displaystyle\times
\sum_{\pi_3}\exp[-i(k_{\pi(1)\lambda_{\pi(1)}},k_{\pi(2)\lambda_{\pi(2)}},
k_{\pi(3)\lambda_{\pi(3)}})], 
\end{array}
\label{eq_c}
\end{equation}
where $k_j$'s are the incoming momenta into the vertex.
We have reversed the signs of variables $\lambda_j$ and $\lambda$, used 
$P_{-\lambda}(-k)=P_{\lambda}(k)$ and the property of the parentheses to 
remove the minus signs in the arguments of the NC phases.

We make a few comments on the above result. First, the NC phases involve only 
on-shell momenta of positive and negative energies. For a given set of 
$\lambda_j$, the permutation sum of NC phases is 
\begin{equation}
\begin{array}{l}
2\cos(k_{1\lambda_1},k_{2\lambda_2},k_{3\lambda_3})+
2\cos(k_{2\lambda_2},k_{3\lambda_3},k_{1\lambda_1})+
2\cos(k_{3\lambda_3},k_{1\lambda_1},k_{2\lambda_2}).
\end{array}
\end{equation}
Note that the above does not simplify into 
$6\cos(k_{1\lambda_1},k_{2\lambda_2})$ etc. as it does in the naive approach 
where all $k_{j\lambda_j}$ are replaced by $k_j$. The reason is that while 
$\sum_jk_j=0$ is always true this is generally not the case with 
$k_{j\lambda_j}$ where 
$k_j^0$ is replaced by $\lambda_j E_{\bfk_j}$: even if this sum vanishes 
for some configuration of $\lambda_j$ and $\bfk_j$, it cannot vanish 
for all configurations. For the case of identical fields considered here, 
there is even no such configuration at all due to kinematics. Furthermore, 
since 
the NC phases depend on time direction parameters $\lambda_j$ we cannot 
exhaust the sum over $\lambda_j$ by using 
$iP_+(k)+iP_-(k)=i\hat{D}_F(k)$ with 
$i\hat{D}_F(k)=i(k^2-m^2+i\epsilon)^{-1}$ 
being the Feynman propagator. These findings are completely different from the naive 
approach. It is intriguing that such differences occur already at  
tree level in perturbation theory and we thus expect that the whole picture 
of perturbation theory will be altered. 
The differences arise from the fact that we are in general not allowed 
to interchange the order of the time-ordering procedure and the star 
multiplication. Only when $\theta_{0i}=0$, the star multiplication does not 
involve time derivatives and the NC phases in eq. $(\ref{eq_c})$ are 
independent of $\lambda_j$, and then the differences disappear.

The external lines in $\hat{G}$ may be amputated by multiplying by an inverse 
Feynman propagator $(i\hat{D}_F)^{-1}$ for each external line and noting that 
\begin{equation}
\begin{array}{rcl}
P_{\lambda}(k)&=&\hat{D}_F(k)\eta_{\lambda}(k),\\
\eta_{\lambda}(k)&=&\displaystyle 
\frac{1}{2}\left(1+\lambda\frac{k_0}{E_{\bfk}}\right).
\end{array}
\end{equation}
We obtain the 1PI vertex for the above $\hat{G}$, 
\begin{equation}
\begin{array}{rl}
&\hat{\Gamma}(k_1,k_2,k_3)\\ 
=&\displaystyle (2\pi)^4\delta^4(\sum_ik_i)
\sum_{\{\lambda_j\}}\prod_{j=1}^3
\left[\eta_{\lambda_j}(k_j)\right]\\
&\displaystyle\times
\sum_{\pi_3}\exp[-i(k_{\pi(1)\lambda_{\pi(1)}},k_{\pi(2)\lambda_{\pi(2)}},
k_{\pi(3)\lambda_{\pi(3)}})]. 
\end{array}
\end{equation}

\subsection{The two by two scattering through cubic interactions}

To motivate our generalization in the next section, we consider the following 
four point function, 
\begin{equation}
\begin{array}{rl}
&G(x_1,x_2,x_3,x_4)\\
=&\displaystyle\int d^4x_5\int d^4x_6
<0|T\left(\pi_1\pi_2\chi_3\chi_4(\pi\star\sigma\star\pi)_5
(\chi\star\sigma\star\chi)_6\right)|0>,
\end{array}
\label{eq_e}
\end{equation}
which would arise from the Lagrangian of cubic interactions amongst real scalar 
fields, 
\begin{equation}
\begin{array}{rcl}
{\cal L}_{\rm int}&=&\displaystyle 
-g_{\pi}(\pi\star\sigma\star\pi)(x)-g_{\chi}(\chi\star\sigma\star\chi)(x).
\end{array}
\end{equation}
We have deliberately introduced nonidentical fields to avoid unnecessary 
complications due to many possible contractions amongst factors of identical 
fields, which just amounts to proper symmetrization of the NC phases as we saw 
in the above example. Our goal will be the S-matrix element of the two by two 
scattering $\pi\pi\to\chi\chi$ and its crossed channels.

It is clear that we should first contract the two $\sigma$ fields. We have, 
for $x_5^0>x_6^0$, 
\begin{equation}
\begin{array}{rl}
&<0|T\left(\pi_1\pi_2\chi_3\chi_4(\pi\star\sigma\star\pi)_5
(\chi\star\sigma\star\chi)_6\right)|0>\\
=&\displaystyle
\pspace_{\bfp}
<0|\cdots(\pi\star e^{-ip_+\cdot x_5}\star\pi)_5
\cdots(\chi\star e^{+ip_+\cdot x_6}\star\chi)_6\cdots|0>, 
\end{array}
\end{equation}
and for $x_5^0<x_6^0$, 
\begin{equation}
\begin{array}{rl}
&<0|T\left(\pi_1\pi_2\chi_3\chi_4(\pi\star\sigma\star\pi)_5
(\chi\star\sigma\star\chi)_6\right)|0>\\
=&\displaystyle
\pspace_{\bfp}
<0|\cdots(\chi\star e^{-ip_+\cdot x_6}\star\chi)_6
\cdots(\pi\star e^{+ip_+\cdot x_5}\star\pi)_5\cdots|0>,
\end{array}
\end{equation}
where the dots represent other fields appropriate to the time ordering and 
$p$ refers to the $\sigma$ field, especially 
$p_{\pm}^0=\pm E_{\bfp}=\pm\sqrt{\bfp^2+m_{\sigma}^2}$. 

Next we consider contractions of $\pi$ and $\chi$ fields. Since 
$[\pi_i,\chi_j]=0$, 
the relative order of $\pi$ and $\chi$ fields is irrelevant; what is relevant 
is the order within the groups $(x_1^0,x_2^0,x_5^0)$ and $(x_3^0,x_4^0,x_6^0)$ 
respectively. Corresponding to $x_5^0>x_6^0$ and $x_6^0>x_5^0$, we have two 
possibilities, $T_{125}T_{346}$ and $T_{346}T_{125}$. Let us study the $\pi$ 
field contraction, 
\begin{equation}
\begin{array}{l}
<0|\cdots T(\pi_1\pi_2\pi_5\star e^{\pm ip_+\cdot x_5}\star\pi_5)\cdots|0>.
\end{array}
\end{equation}
There are $3!$ orders. For example, for $x_1^0>x_2^0>x_5^0$, the above becomes, 
up to disconnected terms, 
\begin{equation}
\begin{array}{rl}
&<0|\cdots \pi_1^+\pi_2^+\pi_5^-\star e^{\pm ip_+\cdot x_5}
\star\pi_5^-\cdots|0>\\ 
=&<0|\cdots(D_{15}\pi_2^++D_{25}\pi_1^+)\star e^{\pm ip_+\cdot x_5}
\star\pi_5^-\cdots|0>\\
=&(D_{15}\star e^{\pm ip_+\cdot x_5}\star D_{25}+(1\leftrightarrow 2))
<0|\cdots|0>,
\end{array}
\end{equation}
where $D_{15}$ and $D_{25}$ refer to the $\pi$ field and $\star$ refers to 
$x_5$. The above is symmetric in $x_1$ and $x_2$ and thus actually corresponds 
to the time order specified by $\tau_{15}\tau_{25}$. The other time orders can 
be similarly computed. Their sum gives the complete result for all orders, 
\begin{equation}
\begin{array}{rl}
&<0|\cdots T(\pi_1\pi_2\pi_5\star e^{\pm ip\cdot x_5}\star\pi_5)\cdots|0>\\ 
=&\displaystyle 
\sum_{\lambda_1\lambda_2}\tau_{15}^{\lambda_1}\tau_{25}^{\lambda_2}
(D_{15}^{\lambda_1}\star e^{\pm ip_+\cdot x_5}\star D_{25}^{\lambda_2}
+(1\leftrightarrow 2))<0|\cdots|0>.
\end{array}
\end{equation}
The last factor in the above is precisely the one for the 
$\chi$ field contraction and is similarly computed. We thus have, 
\begin{equation}
\begin{array}{rl}
&<0|T\left(\pi_1\pi_2\chi_3\chi_4(\pi\star\sigma\star\pi)_5
(\chi\star\sigma\star\chi)_6\right)|0>\\ 
=&\displaystyle\pspace_{\bfp}
\tau_{56}\sum_{\{\lambda_j\}}
\tau_{15}^{\lambda_1}\tau_{25}^{\lambda_2}
\tau_{36}^{\lambda_3}\tau_{46}^{\lambda_4}\\
&\displaystyle\times
\left[D_{15}^{\lambda_1}\star e^{-ip_+\cdot x_5}\star D_{25}^{\lambda_2}
+(1\leftrightarrow 2)\right]
\left[D_{36}^{\lambda_3}\star e^{+ip_+\cdot x_6}\star D_{46}^{\lambda_4}
+(3\leftrightarrow 4)\right]\\
&
+({\rm ~same~as~above~except~}\tau_{56}\to\tau_{65},
~x_{5,6}\to -x_{5,6})\\
=&\displaystyle\pspace_{\bfp}\sum_{\{\lambda_j\}}\sum_{\lambda}
\tau_{15}^{\lambda_1}\tau_{25}^{\lambda_2}
\tau_{36}^{\lambda_3}\tau_{46}^{\lambda_4}\tau_{56}^{\lambda}\\
&\displaystyle\times
\left[D_{15}^{\lambda_1}\star e^{-ip_{\lambda}\cdot x_5}\star 
D_{25}^{\lambda_2}+(1\leftrightarrow 2)\right]
\left[D_{36}^{\lambda_3}\star e^{+ip_{\lambda}\cdot x_6}\star 
D_{46}^{\lambda_4}+(3\leftrightarrow 4)\right], 
\end{array}
\end{equation}
where $D_{36}^{\lambda_3}$ and $D_{46}^{\lambda_4}$ refer to the $\chi$ 
field and the star in the second factor is with respect to $x_6$. In the 
second equality we have made the shift $\bfp\to -\bfp$ for $\lambda=-$.

Using the representations as shown in eq. $(\ref{eq_rep})$ and the same 
trick that led to eq. $(\ref{eq_b})$, we make the above Green 
function ready for transformation into momentum space. For example, 
\begin{equation}
\begin{array}{rl}
&\tau_{15}^{\lambda_1}\tau_{25}^{\lambda_2}
\tau_{36}^{\lambda_3}\tau_{46}^{\lambda_4}\tau_{56}^{\lambda}
\left[D_{15}^{\lambda_1}\star e^{-ip_{\lambda}\cdot x_5}\star 
D_{25}^{\lambda_2}\right]
\left[D_{36}^{\lambda_3}\star e^{+ip_{\lambda}\cdot x_6}\star 
D_{46}^{\lambda_4}\right]\\
=&\displaystyle
\prod_{j=1}^4\left[\int\frac{d^4p_j}{(2\pi)^4}iP_{\lambda_j}(p_j)\right]
\int\frac{d^4p}{(2\pi)^4}iP_{\lambda}(p)\\
&\displaystyle\times
e^{-ip_1\cdot(x_1-x_5)}e^{-ip_2\cdot(x_2-x_5)}
e^{-ip_3\cdot(x_3-x_6)}e^{-ip_4\cdot(x_4-x_6)}
e^{-ip\cdot(x_5-x_6)}\\
&\displaystyle\times
\exp[-i(p_{1\lambda_1},-p_{\lambda},p_{2\lambda_2})]
\exp[-i(p_{3\lambda_3},+p_{\lambda},p_{4\lambda_4})]. 
\end{array}
\end{equation}
We can now integrate over $x_5$ and $x_6$, which results in two factors 
of $\delta$ functions, then transform into momentum space, and sum over 
all terms, 
\begin{equation}
\begin{array}{rl}
&\hat{G}(k_1,k_2,k_3,k_4)\\
=&\displaystyle
\prod_{j=1}^4\left[\int d^4x_j e^{-ik_j\cdot x_j}\right]
G(x_1,x_2,x_3,x_4)\\
=&\displaystyle 
(2\pi)^4\delta^4(\sum_i k_i)\sum_{\{\lambda_j\}}\sum_{\lambda}
\prod_j\left[iP_{\lambda_j}(k_j)\right]iP_{\lambda}(p)\\
&\displaystyle\times 
\left(\exp[-i(k_{1\lambda_1},-p_{\lambda},k_{2\lambda_2})]
+(1\leftrightarrow 2)\right)
\left(\exp[-i(k_{3\lambda_3},+p_{\lambda},k_{4\lambda_4})]
+(3\leftrightarrow 4)\right)\\ 
=&\displaystyle 
(2\pi)^4\delta^4(\sum_i k_i)\sum_{\{\lambda_j\}}\sum_{\lambda}
\prod_j\left[iP_{\lambda_j}(k_j)\right]iP_{\lambda}(p)\\
&\displaystyle\times 
2\cos(k_{1\lambda_1},-p_{\lambda},k_{2\lambda_2})~
2\cos(k_{3\lambda_3},+p_{\lambda},k_{4\lambda_4}), 
\end{array}
\end{equation}
with $p=k_1+k_2=-k_3-k_4$. Be careful that 
$k_{1\lambda_1}+k_{2\lambda_2}\ne p_{\lambda}\ne 
-k_{3\lambda_3}-k_{4\lambda_4}$. 
For comparison, in the naive approach the above would be 
\begin{equation}
\begin{array}{rcl}
(2\pi)^4\delta^4(\sum_i k_i)\prod_j[i\hat{D}_F(k_j)]i\hat{D}_F(p)
~2\cos(k_1,k_2)~2\cos(k_3,k_4), 
\end{array}
\end{equation}
which according to our preceding analysis is correct only for 
$\theta_{0i}=0$. The 1PI function is obtained by amputation,  
\begin{equation}
\begin{array}{rl}
&\hat{\Gamma}(k_1,k_2,k_3,k_4)\\
=&\displaystyle 
(2\pi)^4\delta^4(\sum_i k_i)\sum_{\{\lambda_j\}}\sum_{\lambda}
\prod_j\left[\eta_{\lambda_j}(k_j)\right]iP_{\lambda}(p)\\
&\displaystyle\times 
2\cos(k_{1\lambda_1},-p_{\lambda},k_{2\lambda_2})~
2\cos(k_{3\lambda_3},+p_{\lambda},k_{4\lambda_4}). 
\end{array}
\end{equation}

Let us now extract the S-matrix element for on-shell particles from 
the above 1PI function. We take the example of $\pi\pi\to\chi\chi$ 
scattering. This means, 
$k_1^0=+E_{\bfk_1},k_2^0=+E_{\bfk_2},k_3^0=-E_{\bfk_3},
k_4^0=-E_{\bfk_4}$. Thus only one term in the sum over 
$\{{\lambda_j}\}$ contributes due to $\eta_-(k_1)=0$ etc. Including 
the coupling factors as well, the transition amplitude is 
\begin{equation}
\begin{array}{rl}
&iT\left(\pi(k_1)+\pi(k_2)\to\chi(k_3)+\chi(k_4)\right)\\
=&\displaystyle 
(2\pi)^4\delta^4(k_1+k_2-k_3-k_4)\sum_{\lambda}
iP_{\lambda}(p)(-ig_{\pi})(-ig_{\chi})\\
&\displaystyle\times 
2\cos(k_{1+},-p_{\lambda},k_{2+})~
2\cos(-k_{3+},p_{\lambda},-k_{4+})\\
=&\displaystyle
(2\pi)^4\delta^4(k_1+k_2-k_3-k_4)i{\cal A},  
\end{array}
\end{equation}
where $p=k_1+k_2=k_3+k_4$ and 
${\cal A}$ is the amplitude with the usual normalization as 
computed from Feynman diagrams in ordinary quantum field theory.
Note also that we have reversed the signs of $k_3$ and $k_4$ so that 
$k_{3-}\to -k_{3+},k_{4-}\to -k_{4+}$. 
More explicitly, 
\begin{equation}
\begin{array}{rl}
&T\left(\pi(k_1)+\pi(k_2)\to\chi(k_3)+\chi(k_4)\right)\\
=&\displaystyle
-g_{\pi}g_{\chi}(2\pi)^4\delta^4(k_1+k_2-k_3-k_4)\\
&\displaystyle\times\left[
\frac{2\cos(k_{1+},-p_+,k_{2+})2\cos(-k_{3+},p_+,-k_{4+})}
{2E_{{\bf k}_1+{\bf k}_1}(k_1^0+k_2^0-E_{{\bf k}_1+{\bf k}_1}+i\epsilon)}
\right.\\
&\displaystyle+\left.
\frac{2\cos(-k_{3+},p_-,-k_{4+})2\cos(k_{1+},-p_-,k_{2+})}
{2E_{{\bf k}_3+{\bf k}_4}(-k_3^0-k_4^0-E_{{\bf k}_3+{\bf k}_4}+i\epsilon)}
\right].
\end{array}
\label{eq_d}
\end{equation}

We make a few remarks concerning the S-matrix calculation. First, the 
crossed channels of the above process may be obtained similarly. For 
example, for $\pi\chi\to\pi\chi$ scattering, we may choose 
$k_1^0=+E_{\bfk_1},k_2^0=-E_{\bfk_2},k_3^0=+E_{\bfk_3},
k_4^0=-E_{\bfk_4}$. Second, from the above detailed analysis it is clear 
how to calculate the most complicated case of identical particle 
scattering through their self-interactions. We should include all 
possible Feynman diagrams and for each of them employ the same analysis 
which just amounts to more symmetrization at the vertices with respect to 
identical fields. In this way we get the following contributions to the 
amputated four-point Green function of the $\pi$ field at the lowest 
level in ${\cal L}_{\rm int}=-g_{\pi}\pi\star\pi\star\pi$, 
\begin{equation}
\begin{array}{rl}
&\hat{\Gamma}(k_1,k_2,k_3,k_4)\\
=&\displaystyle 
-g_{\pi}^2(2\pi)^4\delta^4(\sum_ik_i)\sum_{\{\lambda_j\}}\sum_{\lambda}
\prod_{j=1}^4\left[\eta_{\lambda_j}(k_j)\right]\left(A_s+A_t+A_u\right), 
\end{array}
\end{equation}
where $A_s,A_t,A_u$ are from $s-,t-,u-$channels respectively, 
\begin{equation}
\begin{array}{rcl}
A_s&=&\displaystyle iP_{\lambda}(p_s)
\sum_{\pi_3}\exp[-i(k_{1\lambda_1},k_{2\lambda_2},-p_{s\lambda})]
\sum_{\pi_3}\exp[-i(k_{3\lambda_3},k_{4\lambda_4},+p_{s\lambda})],\\
A_t&=&\displaystyle iP_{\lambda}(p_t)
\sum_{\pi_3}\exp[-i(k_{1\lambda_1},k_{3\lambda_3},-p_{t\lambda})]
\sum_{\pi_3}\exp[-i(k_{2\lambda_2},k_{4\lambda_4},+p_{t\lambda})],\\
A_u&=&\displaystyle iP_{\lambda}(p_u)
\sum_{\pi_3}\exp[-i(k_{1\lambda_1},k_{4\lambda_4},-p_{u\lambda})]
\sum_{\pi_3}\exp[-i(k_{3\lambda_3},k_{2\lambda_2},+p_{u\lambda})],
\end{array}
\end{equation}
with $p_s=k_1+k_2,p_t=k_1+k_3,p_u=k_1+k_4$. Here $\pi_3$ refers to the 
$3!$ permutations of the three momenta appearing in each factor of the 
NC phase sums. It is straightforward to project the S-matrix element 
from the above which we shall not write down. And there is also no 
problem to extend to more complicated interactions like $\varphi^4$. 
Finally, if our aim is restricted to the S-matrix for on-shell 
particles, we may proceed more directly from the expectation values of 
the S-operator. For the above example, we need compute the following 
quantity, 
\begin{equation}
\begin{array}{rl}
&<\chi\chi|S|\pi\pi>\\
=&\displaystyle
(-ig_{\pi})(-ig_{\chi})\int d^4x_5\int d^4x_6\\
&\displaystyle\times<\chi(k_{3+})\chi(k_{4+})|
T\left((\pi\star\sigma\star\pi)_5
(\chi\star\sigma\star\chi)_6\right)
|\pi(k_{1+})\pi(k_{2+})>\\
\displaystyle
&+{\rm ~higher~orders},
\end{array}
\end{equation}
which just corresponds to a special assignment of the time order in 
eq. $(\ref{eq_e})$, namely $\pi_1$ and $\pi_2$ in the far past, 
$\chi_3$ and $\chi_4$ in the far future, and others in between. This 
is precisely the contributing part in the above analysis. It is thus 
no doubt that the results for the S-matrix coincide. The main 
advantage of coping with the Green function for this purpose is that 
we may project all physical processes from the same Green function.

\section{Generalization}

The structure shown in eq. $(\ref{eq_d})$ looks familiar to us and is 
very suggestive. Actually it is nothing but the `` old-fashioned '', 
time-ordered perturbation theory 
$\cite{schweber}\cite{sterman}$ 
modified properly to NC field theory. 
This fits also on a somewhat more formal level to the pragmatic
point of view which we assume here. The S-operator maps on the one hand
any prepared incoming state onto the respective outgoing one, but this
can also be considered as the time tranport of this incoming state into
the outgoing one. As long as we can represent it as a time-ordered 
exponential with a Hermitian exponent times the pure imaginary unit $i$
we have formal unitarity and thus satisfy the first requirement for 
true unitarity which means conservation of the transition probability 
in the sense of quantum mechanics -- to which we shall return  
in a separate publication. 

The time-ordered Feynman diagrams for the above example are depicted 
in Fig. $1$. In the language of the time-ordered perturbation theory, 
a physical process is virtualized as a series of transitions between 
physical intermediate states that are sequential in time. The transition 
amplitude is weighted by the interaction vertices which are evaluated 
for on-shell momenta if they depend on them and by the energy deficit 
of the intermediate states. Realizing this, it becomes obvious how to 
proceed in the general case. For further applications, we give below the 
prescriptions for computing the on-shell transition matrix $T$ at some 
fixed order in perturbation, which are readily generalized from the 
ordinary ones $\cite{schweber}\cite{sterman}$. The additional piece for 
general off-shell amputated Green functions will be described later on.

\begin{center}
\begin{picture}(200,150)(0,0)
\SetOffset(0,30)
\ArrowLine(0,0)(30,40)\ArrowLine(60,0)(30,40)
\ArrowLine(30,40)(30,80)
\ArrowLine(30,80)(0,120)\ArrowLine(30,80)(60,120)
\Text(8,25)[]{$k_1$}\Text(52,25)[]{$k_2$}
\Text(8,95)[]{$k_3$}\Text(52,95)[]{$k_4$}
\Text(40,60)[]{$p$}
\Text(30,-10)[]{$(a)$}

\SetOffset(100,30)
\ArrowLine(0,0)(20,80)\ArrowLine(40,0)(20,80)
\ArrowLine(20,80)(60,40)
\ArrowLine(60,40)(40,120)\ArrowLine(60,40)(80,120)
\Text(3,40)[]{$k_1$}\Text(37,40)[]{$k_2$}
\Text(43,80)[]{$k_3$}\Text(77,80)[]{$k_4$}
\Text(40,52)[]{$p$}
\Text(40,-10)[]{$(b)$}

\SetOffset(0,0)
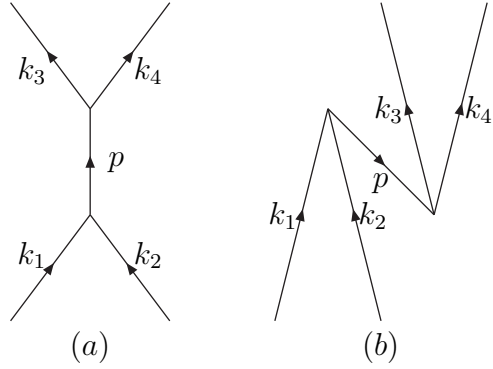
\Text(100,0)[]{Fig. 1: Time-ordered diagrams corresponding to eq. 
$(\ref{eq_d})$. Time flows upwards.}
\end{picture}
\end{center}

\begin{enumerate}
\item{Draw all Feynman diagrams for the process under consideration. 
For each Feynman diagram draw all of its time-ordered diagrams. 
Only the time order of interacting vertices is relevant and 
indistinguishable time orders are counted only once. Each time-ordered 
diagram is computed by putting together the following factors.}

\item{Associate with each internal line (with a spatial momentum $\bfk$) 
a phase space integral $\displaystyle\pspace_{\bfk}$.}

\item{Associate with each vertex $v$, which is formed by internal lines 
$j$ and external ones $e$ of incoming spatial momenta $\bfq_a$ 
and which has the interaction pattern in the Lagrangian 
$-g\psi_1\star\psi_2\star\cdots\psi_n$, 
an interaction factor 
$g\exp[-i(q_{1\lambda_1},q_{2\lambda_2},\cdots,q_{n\lambda_n})]$.
$\lambda_a=+(-)$ if the vertex $v$ is the later (earlier) end of the 
line $a$. The initial (final) particles $e$ are always counted as earlier 
(later) than the vertex $v$. Symmetrize the above factor with respect to 
identical fields.
Impose spatial momentum conservation at the vertex $v$ by multiplying 
$(2\pi)^3\delta^3(\sum_a\bfq_a)$.}

\item{Associate with each intermediate state occurring between two sequential 
vertices (earlier $v_1$ and later $v_2$) a factor of energy deficit, 
$[\sum_e(\pm p_e^0)-\sum_j E(\bfk_j)+i\epsilon]^{-1}$. 
Here $\sum_e(\pm p_e^0)$ is the algebraic sum of the zero-th components of 
external momenta entering ($+$) or leaving ($-$) the diagram before and 
including the earlier vertex $v_1$. $E(\bfk_j)$ is the on-shell positive 
energy of the $j$-th line contained in this intermediate state.}

\item{Multiply by a global factor of $-2\pi\delta(\sum_e(\pm p_e^0))$ and 
a symmetry factor $1/S$ which excludes that of indistinguishable diagrams 
mentioned above.}
\end{enumerate}

The above prescriptions would be precisely the same as obtained in 
ordinary field theory if we could interprete the vertex factor as a kind of 
numerator arising from spin. In ordinary relativistic field theory we can
recast the time-ordered perturbation theory into a covariant form in 
terms of Feynman diagrams. So, it is tempting to ask why this is not 
possible in NC field theory. Of course, Lorentz invariance is lost at the 
very beginning and it is not guaranteed that a seemingly covariant 
formalism exists and is equivalent to the time-ordered one if it does. 
But this is not the whole point. As far as Feynman diagrams are concerned, 
we can always treat $\theta_{\mu\nu}$ as if it were a Lorentz tensor and 
there will be no problem if we do not use any special reference frame 
for calculation since we could not return back by a transformation 
afterwards $\cite{liao}$. We could also consider $\theta_{\mu\nu}$ as 
some background field and assign to it a transformation law so that the 
above consequence still applies. In the following we present an argument 
that in NC field theory formulated via time-ordered perturbation theory one 
cannot reproduce the seemingly covariant results of the naive approach. Our 
time-ordered version seems however to be a safe starting point as 
far as quantum mechanics still applies to NC spacetime. A key element of it 
is the highly nonlocal character of NC interactions. 

Let us first recall briefly how to shift from the covariant perturbation 
theory to the time-ordered one in ordinary field theory. For a detailed 
account of the topic we refer the interested reader to 
Ref. $\cite{sterman}$ 
for a nice presentation. For this purpose, we consider the contribution 
from a Feynman diagram to a general, unamputated and connected Green 
function in momentum space. One first expresses the $\delta$ function of 
the zero-th components of 
momenta at each vertex in terms of a time integral. Then, one rearranges 
the product of time integrals thus obtained in a time-ordered way. This 
is followed by integrating over the zero-th momentum components, which is 
typically of the following form, 
\begin{equation}
\begin{array}{rl}
&\displaystyle\int_{-\infty}^{\infty} 
dk_0\frac{i}{k^2-m^2+i\epsilon}e^{-ik_0 t}f(k_0)\\
=&\displaystyle\int_{-\infty}^{\infty} 
dk_0\frac{i}{(k_0-E_{\bfk}+i\epsilon)(k_0+E_{\bfk}-i\epsilon)}
e^{-ik_0 t}f(k_0). 
\end{array}
\end{equation}
Here $t$ is the difference of the time variables introduced above between 
the two vertices connected by the line carrying momentum $k_{\mu}$ in the 
same direction of $t$. $f(k_0)$ is usually a polynomial of finite degree 
and analytic in the complex $k_0$ plane. The above integral is 
evaluated using contours. For $t>(<)0$, one closes the contour in the 
lower (upper) half plane picking up the residue at 
$k_0=E_{\bfk}-i\epsilon$ ($k_0=-E_{\bfk}+i\epsilon$) with the result, 
\begin{equation}
\begin{array}{l}
\displaystyle 
\frac{\pi}{E_{\bfk}}\left[
\tau(t)e^{-i(E_{\bfk}-i\epsilon)t}f(+E_{\bfk})+
\tau(-t)e^{+i(E_{\bfk}-i\epsilon)t}f(-E_{\bfk})\right]. 
\end{array}
\end{equation}
A crucial condition for the above manipulation is that $f(k_0)$ must 
not blow up at the lower (upper) infinite semicirle in the complex plane 
faster than $e^{-ik_0t}$ decays. Finally, one completes the time integrals 
sequentially and arrives at the result in the time-ordered perturbation 
theory. Now let us try to do the opposite in NC theory from the 
time-ordered perturbation theory to the covariant one by turning around 
the above procedure. In this case the function $f(\pm E_{\bfk})$ is an 
NC phase which is essentially an exponential (superposition) of the form 
$\exp(\pm iE_{\bfk}\tilde{k}^{\prime 0})$, where 
$\tilde{k}^{\prime 0}=\theta_{0i}k^{\prime i}$ with $\bfk^{\prime}$ being 
a spatial momentum of some other internal or external line. If the above 
procedure were reversible, the corresponding function in the complex $k_0$ 
plane would be something like $\exp(ik_0\tilde{k}^{\prime 0})$ so that 
the naive result might have a chance to be recovered. But this 
is impossible because it is not guaranteed that it increases slower 
than $e^{-ik_0t}$ decays. Actually whether it decays or blows up depends 
on the sign of $\tilde{k}^{\prime 0}$ which itself changes with 
$\bfk^{\prime}$. This thus interferes with the above contour integration. 
One may argue that we may shift $t$ to absorb $\tilde{k}^{\prime 0}$. 
This is again not legitimate since $t$ is a time difference and doing so 
simply spoils the time-ordering procedure which is a key bridge to relate 
the two formalisms. Furthermore, a connected diagram has certainly more 
than one line; the above shift of $t$'s, if it worked at all for one of 
them, would also interfere with each other making the trick totally 
useless. The above argument fails only if at tree level 
$\tilde{k}^{\prime 0}$ is an external momentum and happens to vanish. 
But this is a very special kinematic configuraton if possible at all. 
We surely cannot rely on this in favour of the 
naive approach. From this analysis it is also clear that the main obstacle 
originates from the nonlocal exponential interaction that is intrinsic in 
NC field theory. 

Finally we extend the above prescriptions to amputated and connected Green 
functions by adding the following rule concerning external lines for 
time-ordered diagrams. It should also be applied to each individual 
diagram in which the connection of external lines to vertices is fixed. 

\begin{description}
\item{6. Multiply by a factor of $\eta_{\lambda_e}(p_e)$ for each external 
line with incoming momentum $p_e$ and time direction parameter 
$\lambda_e$ which is $+(-)$ if it connects to a(n) later (earlier) vertex.  
This same $\lambda_e$ also appears in the preceding vertex factors (where 
it takes one of the signs for S-matrix). Sum over the set $\{\lambda_e\}$.} 
\end{description}

\section{Discussion and conclusion}

Based on the assumption that a time-ordered expansion of a formally 
unitary time evolution operator is a good starting point also on NC 
spacetime, we studied perturbative NC field theory which turns out 
to be the time-ordered perturbation theory adapted properly to the 
NC case. This was achieved by a detailed analysis of some 
exemplifying Green functions which we worked out with care, and then 
extended to the general case. We found no obstacles in implementing 
noncommutativity in perturbation theory whether time commutes with 
space or not. We provided prescriptions for computing S-matrix 
elements and amputated Green functions. 

Although we only treated scalar fields in this paper, we expect no 
problems with spinor fields as it is the case on ordinary spacetime. 
Since the spinor effect amounts to an additional numerator associated 
with a propagator, it plays a similar role as a vertex, namely, the 
momentum contained therein will be on-shell with positive or negative 
energy. The situation is more complicated for gauge bosons. But again 
as in the usual theory there should be no problem at least in the 't 
Hooft-Feynman gauge. Our method also applies to any dimensions.

The NC perturbation theory thus obtained is already different at tree 
level from the naive approach followed in the literature. The interaction 
vertices involve only on-shell momenta of positive or negative energy 
of participating particles. The basic quantity connecting vertices is 
not the causal Feynman propagator but the individual propagation 
functions of positive and nagative frequency. These elements are naturally 
incorporated in the framework of the time-ordered perturbation theory. In 
contrast to the ordinary field theory, it seems impossible to recast the 
NC time-ordered theory into a covariant form as has been assumed in the 
naive approach. We attributed this difference to the highly nonlocal 
character of phase-type NC interactions which has significant impact on 
the analyticity properties of Green functions in the complex energy plane. 

Since the whole picture for perturbation theory has been changed, we expect 
some of the phenomena found previously in the naive approach will also be 
altered. Amongst them we would like to mention briefly the unitarity issue 
which we will detail soon in a separate paper. Since the NC phase now 
involves only on-shell momentum, it is independent of the zero-th component 
of a generally off-shell four-momentum. The analyticity properties of Green 
functions in the complex plane of the zero-th component will thus be very 
different from that in the naive approach. Furthermore, the right-hand side 
of the unitarity relation for Green functions is also modified due to the 
change of vertices, which does not seem to have been noticed thus far. 
Considering all of this it is quite reasonable to expect that unitarity 
will be practically preserved as it is formally built in the time-ordered 
perturbation theory. This may have some hints for the interplay of string 
and field theory.

{\bf Acknowledgements}

Y.L. would like to thank M. Chaichian for a visit at the Helsinki 
Institute of Physics and its members for hospitality. He enjoyed many 
encouraging discussions with M. Chaichian, P. Presnajder and A. Tureanu.
K.S. is grateful to D. Bahns and K. Fredenhagen for clarifying 
discussions on their work.

As we were preparing the manuscript a new preprint $\cite{rim}$ appeared 
in which the ideas developped in Ref. $\cite{bahns}$ were further 
elaborated on by some examples.

\end{document}